\documentclass{llncs}

\usepackage{amssymb}
\usepackage{graphicx}

\usepackage{subfigure}

\usepackage{url}

\usepackage[utf8]{inputenc}

\usepackage[turkish]{babel}

\begin{document}

\mainmatter  

\title{Enerji İzleme Yazılımları için Merkezi ve Genel bir Mimari}
\titlerunning{Enerji İzleme Yazılımları için Merkezi ve Genel bir Mimari}
\author{Dilek Küçük, Turan Demirci}
\authorrunning{Küçük ve Demirci}

\institute{TÜBİTAK MAM Enerji Enstitüsü\\
Ankara, Türkiye\\
\email{\{dilek.kucuk,turan.demirci\}@tubitak.gov.tr}}

\maketitle

\begin{abstract}
Enerji alanında değişik yeteneklere sahip birçok yazılım sistemine ihtiyaç duyulmakta ve bu ihtiyaçlara yönelik sistemler geliştirilmektedir. Söz konusu sistemler arasında enerji izleme, bilgi, geniş alan izleme ve kontrol sistemleri ile SCADA sistemleri sayılabilir. Bu sistemlerin en önemli ve yaygınlarından biri enerji izleme yazılımlarıdır. Bu çalışmada, enerji alanındaki yazılım sistemleri kısaca gözden geçirildikten sonra özellikle enerji izleme yazılımları için merkezi ve genel bir yazılım mimarisi sunulmuştur. Ardından, bu mimariye dayalı izleme yazılımlarının gerçekleştirildiği geniş ölçekli projelerden örnekler verilmiştir. Bu çalışma-nın, enerji alanında yazılım geliştirme projeleri için önemli bir kaynak olacağı düşünülmektedir.\\

\textbf{Anahtar kelimeler:} sistem mimarisi, yazılım mimarisi, enerji izleme yazılımları, enerji yazılım sistemleri.\\

\textbf{Abstract} There is need for several software systems within the energy domain and corresponding systems are being developed to satisfy these needs. These systems include energy monitoring, information, wide area monitoring and control systems, and SCADA systems. Energy monitoring systems are one of the most important and common systems among them. In this study, after briefly reviewing several of the software systems within the energy domain, a centralized and generic software architecture for energy monitoring systems is presented. Next, sample projects are described in which energy monitoring systems based on this architecture have been implemented. We envisage that this study will be an important resource for software projects in the energy domain.\\

\textbf{Keywords:} system architecture, software architecture, energy monitoring software, energy software systems.

\end{abstract}

\section{Giriş}\label{giris}
Enerji, hem toplumdaki her bir bireyi hem de daha geniş ölçekte endüstriyel kuruluşları önemli oranda etkileyen bir konudur. Bu nedenle; enerji verilerinin uygun şekilde toplanması, çözümlenmesi, saklanması, sunulması ve kontrol edilmesi de oldukça fazla önem arz etmektedir. Bu amaçlarla, enerji alanında çoğunlukla geniş ölçekli olmak üzere farklı yeteneklere sahip yazılımların geliştirilmesi ihtiyacı ortaya çıkmakta ve bu yönde geliştirmeler yapılmaktadır. Enerji alanında yaygın kullanılan bazı yazılımlar şunlardır: izleme sistemleri, bilgi sistemleri, geniş alan izleme ve kontrol sistemleri, SCADA sistemleri ve enerji yönetim sistemleri, karar destek sistemleri, ve tahmin sistemleri.

Bu çalışmamızın konusunu oluşturan enerji izleme yazılımları; enerji verisinin toplandığı (otomatik) ölçüm yazılımlarını, bunların merkezi bir veri tabanında saklanması için gerekli haberleşme yazılımlarını, veri tabanı bileşenini ve son olarak da saklanan verilerin ilgili kullanıcılara sunulduğu kullanıcı arayüzü uygulamalarını kapsamaktadır. Bu çalışmada, enerji izleme yazılımları için genel ve merkezi bir sistem mimarisi sunulmuştur. Ayrıca, bu mimarinin kullanıldığı tamamlanmış ve kullanımda olan geniş ölçekli projeler de tanıtılmıştır. Çalışmamızın, enerji alanından yazılım geliştirilecek projeler için önemli bir kaynak olacağı öngörülmektedir. Bildirimizin devamı şu şekilde yapılandırılmıştır: 2. bölümde, enerji alanın-daki örnek yazılım sistemleri hakkında genel bilgiler verilmiştir; 3. bölümde enerji izleme yazılımları için önerdiğimiz genel ve merkezi yazılım mimarimiz, bu mimarinin kullanıldığı örnek projelerle birlikte tanıtılmıştır; son olarak 4.bölümde de çalışmamızın sonuçları verilmiştir.

\section{Enerji Alanında Yazılım Örnekleri}\label{bilgi}
Elektrik şebekesi {--}çok genel bir ifadeyle{--} enerji üretim, iletim, dağıtım ve tüketim alt sistemlerinden oluşmaktadır \cite{Grigsby2012}. Elektrik üretim sistemi bünyesinde bulunan elektrik santrallerinde üretilen enerji, önce elektrik iletim sistemine iletilmekte, buradan da enerji dağıtım sistemine ulaştırılmaktadır. Elektrik dağıtım sistemi de enerjiyi son kullanıcılar olan ağır sanayi ve sanayi kuruluşlarına ve şehirlere sağlamaktadır. Şebekenin gereken şekilde işletilmesini sağlayabilmek için; uygun noktalarda verileri toplamak, depolamak, izlemek, analiz etmek, yönetimsel karar sürecinde kullanmak önem arz etmektedir. Bu nedenle, enerji alanı için farklı yeteneklere sahip çok sayıda yazılım sistemi geliştirilmektedir. Aşağıda, bu yazılımlardan en yaygın olanlarının kısa açıklamaları verilmiştir:
\vspace{2mm}

\noindent\textbf{\emph{İzleme Sistemleri:}} Elektrik sistemindeki güç ve güç kalitesi parametreleri gibi çok çeşitli parametrelerinin çeşitli çözünürlükte verilerinin farklı ölçüm noktalarından otomatik olarak toplanması, merkezi veya dağıtık yapıda bir veya birden fazla veri tabanında saklanması ve sistem kullanıcılarına uygun arayüzlerle sunulması yeteneklerine sahip olan sistemlerdir \cite{Demirci2011}. Ayrıca, bu bölümde tanıtılan enerji yazılımlarının birçoğu ya kendine has bir izleme modülüne sahiptir yada kendisi bir çeşit izleme yazılımı olarak kabul edilebilir.
\vspace{1mm}

\noindent\textbf{\emph{Bilgi Sistemleri:}} Genellikle, enerji alanında faaliyet gösteren kurum ve kuruluşların kendi yönetimlerindeki ekipmanlar hakkındaki değişmez ve değişken nitelikteki bilgileri takip etmek için kullandıkları sistemlerdir. Enerji bilgi sistemlerinde ilgili ekipmanın marka ve model gibi değişmez bilgilerinin yanı sıra o anki durum (açık/kapalı) ve arıza geçmişi gibi değişken bilgileri de tutulmakta ve takip edilmektedir \cite{Eren2015}.
\vspace{1mm}

\noindent\textbf{\emph{Geniş Alan İzleme ve Kontrol Sistemleri:}} Bu sistemler, kritik noktalarda akım ve gerilim büyüklükleri ile açılarının (fazörlerin) çok yüksek çözünürlüklü (50 ms. veya 100 ms.'lik ortalamalar gibi) olarak hesaplandıktan sonra bir merkezde toplanarak analiz edilmesine olanak sağlayan sistemlerdir \cite{Terzija2011}. Ayrıca bu analizler sonucu otomatik olarak sistemin kontrol edilmesi de bu sistemlerle sağlanabilir. Bu sistemler, genel nitelikleri bakımından izleme sistemi başlığı altında da değerlendirilebilir.
\vspace{1mm}

\noindent\textbf{\emph{SCADA Sistemleri:}} SCADA (Supervisory Control And Data Acquisition) sistemleri, elektrik şebekesinin anlık olarak izlenerek uzaktan kontrol edilmesini sağlayan sistemlerdir \cite{Karnouskos2011}. Bu sistemler de izleme yeteneklerine sahip olmakla birlikte, izledikleri noktalara kontrol sinyalleri de göndererek yerel işlemlerin yapılabilmesine olanak sağlamaktadır. SCADA sistemlerinin, optimizasyon gibi ek yeneteklere sahip sürümlerine genellikle enerji yönetim sistemleri adı verilmektedir.
\vspace{1mm}

\noindent\textbf{\emph{Karar Destek Sistemleri:}} Enerji alanında karar destek sistemleri; elektrik sistemi yöneticilerinin sistem üzerinde uygulanacak kısa ve uzun süreli işletmeyle veya planlamayla kararlarına çeşitli analizler yoluyla destek olan sistemlerdir \cite{Demirci2011,Eren2015}. Karar destek sistemleri yukarıda tanıtılan bilgi ve izleme sistemlerinin de bünyesinde yer alabilir.
\vspace{1mm}

\noindent\textbf{\emph{Tahmin Sistemleri:}} Enerji tahmin sistemleri arasında yenilenebilir ve değişken nitelikte enerji kaynakları olan rüzgar ve güneş enerjisi tahminlerinin yapıldığı sistemler sayılabilir  \cite{Terciyanli2014}. Ayrıca, bir bölge, şehir veya tüm ülke genelinde enerji tüketiminin (yükün) tahmini de enerji üretim planlaması için önemlidir ve bu amaçla yük tahmin sistemleri geliştirilmektedir \cite{khotanzad1997annstlf}.
\vspace{1mm}

\noindent\textbf{\emph{Özel Koruma Sistemleri:}} Bu sistemler; genellikle elektrik şebekesinin bir bölümünde meydana gelen olağandışı bir durumda, önceden tanımlanmış uygun bir çözümü otomatik olarak uygulayarak (ilgili ekipmanı sistemin genelinden ayırmak gibi) bu durumdan dolayı sistemin genelinde meydana gelecek zararı en aza indirmeye çalışan yazılım sistemleridir \cite{Ross2013}.
\vspace{1mm}

\noindent\textbf{\emph{Olay Kaydedici Sistemler:}} Elektrik iletim sisteminde; ani gerilim düşmesi, yükselmesi ve kesintisi gibi olaylar sistemdeki anormal durumların belirtileridir. Bu gibi durumların zamanında tespit edilip sonrasında analiz edilebilmesi için olayları otomatik olarak tespit eden olay kaydedici yazılımlar geliştirilmekte ve elektrik sisteminin değişik noktalarındaki olaylar izlenmektedir \cite{Demirci2011}. Olay kaydedici sistemler, izleme sistemleri genel başlığı altında da değerlendirilebilmektedir.

\section{Enerji İzleme için Merkezi ve Genel Yazılım Mimarisi}\label{mimari}
Bu bölümde, özellikle enerji izleme sistemlerinin geliştirilmesinde kullanılabilecek merkezi ve genel bir yazılım mimarisi sunulmaktadır. Bu mimari, büyük ölçekli projeler içerisinde ilgili yazılımların gerçekleştirimi sırasında kullanılmıştır ve böylelikle uygulanabilirliği de örneklendirilmiştir. Aşağıdaki alt bölümlerin ilkinde bu mimari tanıtılmış, ikinci alt bölümde ise bu mimarinin kullanıldığı projelerden örnekler verilmiştir.

\subsection{Yazılım Mimarisi}\label{mimaridetay}
Merkezi yapıdaki enerji izleme sistemleri için tasarladığımız genel yazılım mimarisi Şekil 1'de verilmiştir.

\begin{figure}
\center \scalebox{0.47}{\includegraphics{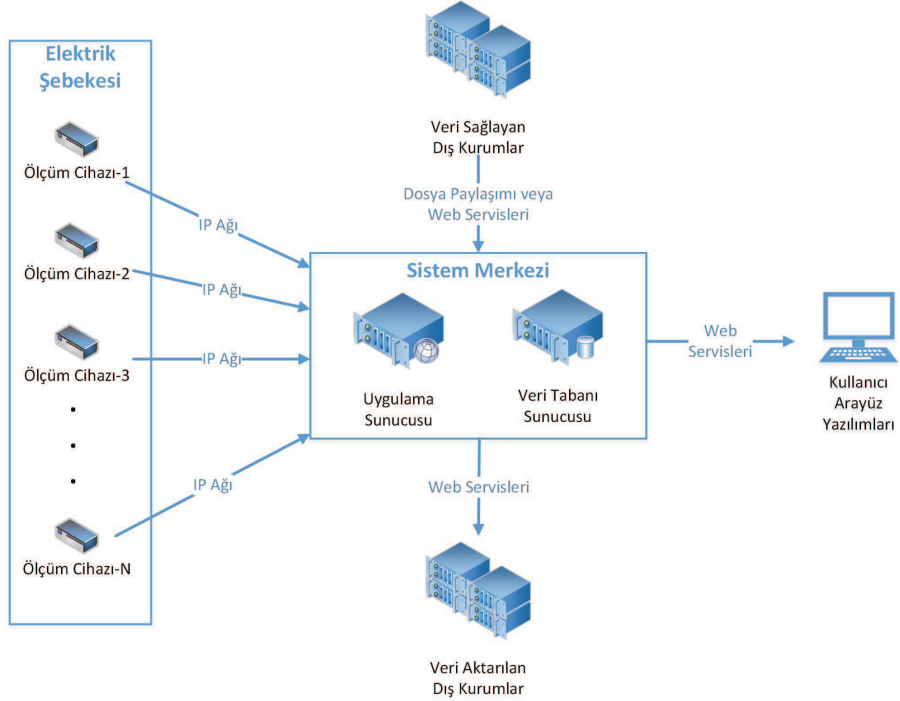}}
\caption{Enerji İzleme Yazılımları için Önerilen Merkezi ve Genel Mimari}
\label{fig:taxonomy}
\end{figure}

Sistem mimarisinde, IP tabanlı bir ağ alt yapısı ile elektrik şebekesinin değişik noktalarına monte edilmiş ölçüm cihazlarından ölçüm verilerini merkezi veri tabanına aktarılmaktadır. Bu aktarımı ilgili haberleşme yazılımları sağlamaktadır. Verilerin güvenli bir şekilde taşınmasının sağlanabilmesi için veri paketleri gönderilmeden önce şifrelenmekte, sunucu tarafında çözülmektedir. Merkezi veri tabanında depolanan ölçüm verileri kullanıcı arayüz yazılımlarına ilgili Web servis altyapısıyla sunulmaktadır. Sistemin cihazlardan otomatik toplanan veriler haricindeki dış kurumlardan veri gereksinimleri büyük boyutlu veriler için dosya paylaşımı yöntemiyle, diğer veriler içinse Web servisleriyle sağlanabilmektedir. Sistemin diğer kurum ve kuruluşlara otomatik veri sağlaması da yine Web servisleriyle sağlanabilmektedir.

Mimarinin merkezi veri tabanında, sürekli bir şekilde değişik çözünürlüklerde (100 ms.'lik, 1-3 sn.'lik, 1-10 dk.'lık ortalamalar gibi) ve uzun süre veri saklandığından, özellikle ölçüm noktası sayısı da fazlaysa, kullanıcı sorgularının dönüş süresi oldukça uzun olabilecektir. Bu nedenle, sistemin kullanıcı sorgularına kabul edilebilir süreler içerisinde cevap dönülmesinin sağlanması amacıyla daha sıklıkla sorgulanması olası olan güncel verilerin bellek üzerinde oluşturulmuş bir veri tabanında, daha eski verilerinse disk üzerinde oluşturulan bir veri tabanında saklanması öngörülmüştür. Bu nedenle mimaride, veri tabanı biri bellek üzerinde, diğer disk üzerinde olmak üzere iki veri tabanından oluşacak şekilde tasarlanmıştır.

\subsection{Örnek Projeler}\label{ornekprojeler}
Aşağıda, tanıttığımız mimariyi temel alarak ilgili izleme sistemlerinin geliştirildiği iki geniş ölçekli proje hakkında bilgiler sunulmuştur:\\

\noindent\textbf{\emph{Güç Kalitesi Milli Projesi:}} Bu projede, Türkiye'nin elektrik iletim sistemindeki önemli noktalarda güç (aktif\slash reaktif\slash görünür güç) ve güç kalitesi parametrelerinin (frekans, harmonikler, kırpışma, dengesizlik vb.) 3 sn.'lik ve 10 dk.'lık ortalamalar şeklinde hesaplanmaları ve bu değerlerin, varsa tespit edilen güç kalitesi olaylarıyla birlikte, merkezi bir veri tabanında tutulması; iletim sistemi operatörü ve diğer sistem kullanıcılarının geliştirilen kullanıcı arayüzleri yoluyla bu parametreleri izleyebilmesi amaçlanmıştır \cite{Demirci2011}. Böylelikle ilgili kullanıcılar iletim sisteminde meydana gelen anormal olayları ve sistemin genel işleyişini izleyerek bu bilgileri karar verme süreçlerinde kullanabilecek duruma gelmiştir. Veri tabanı yönetim sistemi olarak PostgreSQL kullanılmış, ilgili kullanıcı yazılımları Java ile; ölçüm cihazlarındaki analiz yazılımları ile bu cihazların merkezle haberleşmede kullandığı yazılımlar C diliyle geliştirilmiştir. Proje, 2006-2010 yılları arasında tamamlanmıştır ve şu anda iletim sisteminde 800'den fazla ölçüm noktasına ait veriler, proje kapsamında geliştirilen güç kalitesi ölçüm cihazları yoluyla toplanmakta ve izlenmektedir (\url{http://www.guckalitesi.gen.tr/}).

\vspace{2mm}

\noindent\textbf{\emph{Rüzgar Gücü İzleme ve Tahmin Merkezi (RİTM) Projesi:}} Bu projede, Türkiye' deki rüzgar santrallerinin güç üretimleriyle birlikte, santral sahalarındaki rüzgar hızı ve yönü gibi meteorolojik verilerle güç kalitesi parametrelerinin izlenebilmesi, ayrıca bu santraller için çok kısa süreli (6 saate kadar) ve kısa süreli (48 saatlik) rüzgar gücü tahminlerinin yapılması amaçlanmıştır \cite{Terciyanli2014}. Söz konusu proje kapsamında, ölçüm cihazlarından 3 sn.'lik ve 10 dk.'lık ortalamalar olarak güç ve güç kalitesi parametreleri, uygulanabilir santraller için rüzgar ölçüm istasyonların rüzgar hızı ve yönü gibi meteorolojik veriler, santrallerin SCADA sistemlerinden türbinlerin çalışıp çalışmama durumları sürekli olarak merkezi veri tabanlarına aktarılmaktadır. Ayrıca Meteroloji Genel Müdürlüğü, ECMWF ve GFS gibi ulusal ve uluslararası kuruluşlardan dosya paylaşımı yoluyla sürekli olarak orta ölçekli hava tahminleri alınmaktadır. Talep eden ulusal kurumlara da Web servisi yoluyla yetkili oldukları santrallere ait tahminler sağlanmaktadır. Projenin Web arayüzleri JavaServer Faces (JSF) teknolojisi kullanılarak geliştirilmiştir. RİTM projesi 2010-2014 yılları arasında tamamlanmıştır ve şu anda yaklaşık 100 rüzgar santrali proje kapsamında izlenmektedir (\url{http://www.ritm.gov.tr/}).

\section{Sonuç}\label{sonuc}
Elektrik şebekesi bünyesinde enerjinin ölçülmesi, izlenmesi, önceden tahmin edilmesi ve bu bilgiler kullanılarak şebekenin işletilmesi ve planlamalar yapılması özellikle sistem operatörleri için önem arz etmektedir. Bu amaçla bir çok farklı nitelikte yazılım sistemi geliştirilmekte ve kullanılmaktadır. Bunlardan en yaygın olanlarından biri de enerji izleme yazılımlarıdır. Bu çalışmada, merkezi nitelikteki enerji izleme yazılımları için genel bir mimari sunulmuştur. Ardından da bu mimarinin kullanıldığı geniş ölçekli iki proje hakkında bilgiler verilmiştir. Sunduğumuz mimarinin ve proje örneklerinin enerji alanında yazılım geliştirecek araştırmacılar ve yazılım geliştiriciler için faydalı olacağı öngörülmektedir.\\

\noindent \textbf{Teşekkür}: Bu bildiride sunulan çalışma; TÜBİTAK MAM Enerji Enstitüsü Ankara Birimi tarafından yürütülen Güç Kalitesi Milli Projesi (proje no: 106G012) ve Rüzgar Gücü İzleme ve Tahmin Merkezi (RİTM) Projesi (proje no: 5122807) kapsamında yapılmıştır. Yazarlar, Güç Kalitesi Milli Projesi'ni destekleyen TÜBİ-TAK Kamu Araştırmaları Destek Grubu'na (KAMAG) ve RİTM Projesini destekleyen Yenilenebilir Enerji Genel Müdürlüğü'ne (YEGM) teşekkür eder.

\bibliographystyle{splncs}

\end{document}